\documentclass[pra, aps, twocolumn, groupedaddress, showpacs, superscriptaddress]{revtex4}

\usepackage{slashed}
\usepackage{graphicx}
\usepackage{subfigure}
\usepackage[usenames, dvipsnames]{color}
\usepackage{graphics}
\usepackage{hyperref}
\usepackage{bm}
\usepackage{amsfonts}

\usepackage{lipsum}

\hypersetup{backref,
colorlinks=true,
linkcolor=blue,
linktoc=page,
citecolor=blue}

\begin{document}

\title{Spontaneous generation of phononic entanglement in quantum dark-soliton qubits}

\author{Muzzamal I. Shaukat}
\affiliation{ Instituto Superior T\'ecnico, University of Lisbon and Instituto de Telecomunica\c{c}\~{o}es, Torre Norte, Av. Rovisco Pais 1,
Lisbon, Portugal}
\affiliation{CeFEMA, Instituto Superior T\'ecnico, Lisbon, Portugal}
\affiliation{University of Engineering and Technology, Lahore (RCET Campus), Pakistan}
\email{muzzamalshaukat@gmail.com}
\author{Eduardo V. Castro}
\affiliation{CeFEMA, Instituto Superior T\'ecnico, Universidade de Lisboa, Lisboa, Portugal}
\affiliation{Centro de F\'isica das Universidades do Minho e Porto,
Departamento de F\'isica e Astronomia, Faculdade de Ci\'encias,
Universidade do Porto, Porto, Portugal}
\author{Hugo Ter\c{c}as}
\affiliation{Instituto de Plasmas e Fus\~ao Nuclear, Instituto Superior T\'ecnico, Lisboa, Portugal}

\email{hugo.tercas@tecnico.ulisboa.pt}


\begin{abstract}
We show that entanglement between two solitary qubits in quasi one-dimensional Bose-Einstein condensates can be spontaneously generated due to quantum fluctuations. Recently, we have shown that dark solitons are an appealing platform for qubits thanks to their appreciable long lifetime. We investigate the spontaneous generation of entanglement between dark soliton qubits in the dissipative process of spontaneous emission. By driving the qubits with the help of oscillating magnetic field gradients, we observe the formation of long distance steady-state concurrence. Our results suggest that dark-soliton qubits are a good candidates for quantum information protocols based purely on matter-wave phononics. 
\end{abstract}

\maketitle
\section{Introduction}
After the exploitation of entanglement in optical and atomic setups, entanglement generation finds renewed interest in condensed matter systems. Short-distance entanglement has been envisaged for spin or charge degrees of freedom in molecules, nanotubes or quantum dots \cite{Weber2010,Makhlin2001,Franceschi2010, Muzzamal2013, Hanson2007}; owing to the long-range nature of the dipolar ($\sim 1/r^3$) interaction, Rydberg atoms are attractive platforms for large-distance entanglement generation \cite{Gillet2010,Lukin2001,Saelen2011,Urban2009,Hettich2002}. In fact, considerably large separation between atoms is required to transport information at long distances in such systems. To achieve this purpose, a virtual boson mediating the correlation between two qubits is required. Photons are the usual candidate for this task, either for superconducting qubits coupling in the microwave range \cite{Majer2007} or for quantum dots in the visible range \citep{Imamoglu1999,Gallardo2010,Laucht2010}. The investigation to generate two-photon entangled states has been established \cite{Khulud2011}. Plasmons have also been proposed to mediate qubit-qubit entanglement in plasmonic waveguides \cite{Gonzalez2011}. \par 

Thanks to their large coherence times, ultracold gases are natural platforms for quantum information processing, quantum metrology \cite{Sørensen2000}, quantum simulation \cite{Buluta2009}, and quantum computing. In that regard, Bose-Einstein condensates (BECs) have attracted a great deal of interest during the last decades \cite{Davis1995,Anderson1995,Parkins1998}. The macroscopic character of the wavefunction allows BEC to display pure-state entanglement, like in the single-particle case, since all particles occupy the same quantum state. The entanglement between two cavity modes mediated by a BEC has been investigated in Ref. \cite{Ng2009}; two-component BECs have been produced on atom chips with full control of the Bloch sphere and spin squeezing \cite{Böhi2009,Riedel2010}. 

\begin{figure}[t!]
\includegraphics[scale=1]{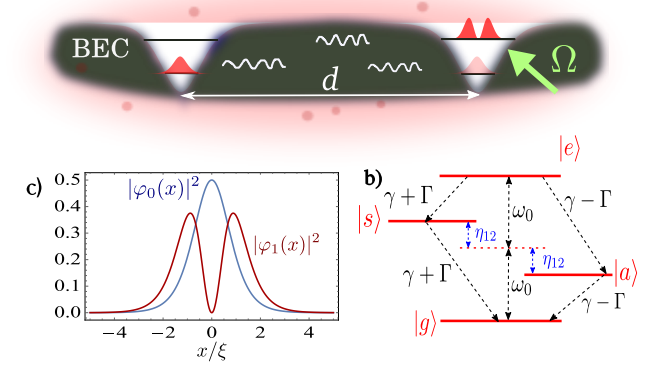}
\caption{(color online) a) Schematic representation of two dark-soliton qubits placed at distance $d$ in a cigar shaped quasi one-dimensional BEC, surrounded by a dilute gas of impurities. b) Collective states of two dark-soliton qubits. Due to the coherent coupling, the two intermediate states $\vert s \rangle$ and $\vert a \rangle$ are maximally entangled. c) Qubit amplitudes in the ground ($\vert \varphi_0(x)\vert^2$) and excited ($\vert \varphi_1(x)\vert^2$) states.}
\label{fig_scheme}
\end{figure}
\par

Another important feature of the macroscopic nature of BECs is the dark-soliton (DS), a structure resulting from the detailed balance between dispersion and nonlinearities. DSs are accompanied by a phase jump, resulting on an extra topological protection \cite{Denschlag, kivshar, burger}. 
The dynamics and stability of DSs in BECs have been a subject of intense research over the last decades \cite{Dutton, Dziarmaga2003,Jackson2007, Mishmash2009}. In that regard, the collision-induced generation of entanglement between uncorrelated quantum solitons has been proposed by Lewenstein et al.  \cite{Lewenstein2009}. The study of collective aspects of soliton gases bring DSs towards applications in many-body physics \cite{gael2005, tercas}. In a recent publication, we have shown that DSs can behave as qubits in quasi one-dimensional (1D) BECs \cite{muzzamal2017}, being excellent candidates to store information given their appreciably long lifetimes ($\sim 100$ ms). {\it Dark-soliton qubits} thus offer an appealing alternative to solid-state and optical platforms, where information processing involves only phononic degrees of freedom: the quantum excitations on top of the BEC state.\par

In this paper, we report on the spontaneous generation of large-distance entanglement between two DS-qubits placed inside a quasi 1D BEC. The entanglement is generated by a combination of the external driving (with the help of magnetic field gradients \cite{navon_2016}) and the quantum fluctuations (phonons) leading to spontaneous and collective emission. We compute the steady-state concurrence for sufficiently large distances, $d\simeq 5\xi/2$, with $\xi$ denoting the healing length, i.e. the size of the soliton core, as depicted in Fig. \ref{fig_scheme}.

The paper is organized as follows: In sec. II, we start with the set of coupled Gross-Pitaevskii and Schr\"odinger equations, to describe the theoretical model based on two DS qubits in a quasi-1D BEC. Here, we also compute the coupling between phonons and DSs. Sec. III describes the effect of  Dicke bases on the spontaneous generation of entanglement. Sec. IV is devoted to the externally driven magnetic field gradient scheme to observe the finite steady-state concurrence, followed by a summary or conclusion in Sec. V.

\section{Theoretical Model}
 We consider two DS placed at a distance $d$ in a quasi 1D BEC. The qubits are formed with the help of an extremely dilute gas surrounding the condensate, whose particles are trapped inside the potential created by the DSs, as illustrated in Fig. \ref{fig_scheme}. At the mean-field level, the system is governed by the Gross-Pitaevskii and the Schr\"odinger equations, respectively describing the BEC and the impurities 
\begin{eqnarray}
i\hbar \frac{\partial \psi}{\partial t}&=&-\frac{\hbar ^{2}}{2m_\psi}\frac{
\partial^{2} \psi}{\partial x^{2}}+g\left\vert \psi\right\vert ^{2}\psi
+\chi\left\vert \varphi\right\vert ^{2}\psi, \nonumber \\
i\hbar \frac{\partial \varphi}{\partial t}&=&-\frac{\hbar ^{2}}{2m_\varphi}\frac{
\partial^{2} \varphi}{\partial x^{2}}+\chi\left\vert \psi _{\rm sol}\right\vert ^{2}\varphi.  \label{gp}
\end{eqnarray}
Here, $\chi$ is the BEC-impurity coupling constant, $g$ is the BEC self-interaction strength, and $m_\psi$ and $m_\varphi$ denote the BEC particle and impurity masses, respectively. The two-soliton profile is \cite{zakharov72, huang}
\begin{equation}
\psi_{\rm sol}(x)=\sqrt{n_0}\prod_{j=1}^2(-1)^{j+1}\tanh \left(\frac{x-x_j}{\xi}\right),
\end{equation}
where $x_{j}=\pm d/2$ are the position of the soliton centroids, $n_{0}$ is the BEC linear density, $\xi =\hbar /\sqrt{g n_{0}m_\psi}$ is the healing length. One possible experimental limitation has to do with inhomogeneities induced by the trap \cite{parker2004}. Fortunately, homogeneous condensates are nowadays experimentally feasible in box-shaped potentials \cite{zoran2013}. This offers additional advantages regarding the scalability (i.e. in a multiple-soliton quantum computer), as uncontrolled phonon mediated soliton-soliton interaction appears when inhomogeneities exist \cite{allen2011}. In this paper, we make our numerical estimates based on homogeneous condensates loaded in box potentials (see Appendix-\ref{Bound states in a dark-soliton potential: dark-soliton qubits}).
\subsection{Quantum fluctuations} 
The total BEC quantum field includes the two-soliton wave function and quantum fluctuations, 
\begin{equation}
\Psi(x)=\psi_{\rm sol}(x)+\sum_j \delta\psi_j(x),
\end{equation}
with $\delta \psi_{j}(x)=\sum_k \left(u^{(j)}_k(x) b_k +v^{*(j)}_k(x)b^{\dagger}_k \right)$ and $b_k$ being the bosonic operators verifying the commutation relation $[b_{k},b^{\dagger}_{q}]=\delta_{k,q}$. The LDA amplitudes $u_k^{(j)}(x)$ and $v_k^{(j)}(x)$ satisfy the normalization condition $\vert u_k^{(j)}(x)\vert ^2 -\vert v_k^{(j)}(x)\vert ^2=1$ and are explicitly given in the Appendix-\ref{Interaction Hamiltonian}.
The total Hamiltonian then reads $H=H_{\rm q}+H_{\rm p}+H_{\rm int}$, where $H_{\rm q}=\sum^{2}_{i=1}\hbar \omega _{0}\sigma _{z} ^{(i)} $ is the qubit Hamiltonian, $\omega _{0}=\hbar(2\nu -1)/(2m_\varphi\xi ^{2})$ is the qubit gap energy, and $\nu=[-1+\sqrt{1+4\chi m_\varphi/gm_\psi}]/2$ is a parameter controlling the number of bound states created by each DS, which operate as qubits (labeled by the states $l=\{0,1\}$) in the range $0.33<\nu <0.80$ (Appendix-\ref{Bound states in a dark-soliton potential: dark-soliton qubits}) \cite{muzzamal2017}. The term $H_{\rm p}=\sum_k \epsilon _{k}b_{k}^{\dagger} b _{k}$ represents the phonon (reservoir) Hamiltonian, where $\epsilon _{k}=\mu \xi \sqrt{k^{2}(\xi^{2}k^{2}+2)}$ is the Bogoliubov spectrum with chemical potential $\mu=gn_{0}$. The interaction Hamiltonian can be constructed as
\begin{equation}
H_{\rm int}=\chi\int dx\Phi ^{\dag }\Psi ^{\dag }\Psi \Phi ,
\label{Int. Ham.}
\end{equation}
where $\Phi(x)=\sum_{l,j} \varphi^{(j)}_{l}(x) a^{(j)}_{l}$ is the impurity field, spanned in terms of boson operators annihilating an impurity in the state (``band") $l$ at site $j$, $a_{l}^{(j)}$. Moreover, $\varphi_0^{(j)}(x)=A_0{\rm sech}^\alpha \left[ (x -x_j)/\xi \right]/\sqrt{2\xi}$ and $\varphi_1^{(j)}(x)=A_1\tanh \left[ (x -x_j)/\xi \right]\varphi_0^{(j)}(x)$ are the Wannier functions relative to Eq. (\ref{gp}), with width $\alpha=\sqrt{\chi m_\varphi/g m_\psi}$ and normalization constants $A_l$ (Appendix-\ref{Interaction Hamiltonian}). Using the rotating wave approximation (RWA), the first-order interaction Hamiltonian, comprising interband terms only, read (Appendix-\ref{Interaction Hamiltonian})%
 \begin{eqnarray}
H_{\rm int} &=&\sum_{k}\sum_{j=1}^{2}\left( g^{(j)}_k\sigma^{(j)} _{+}b_{k}+ g^{(j) \ast}_k \sigma^{(i)} _{-}b_{k}^{\dag }\right)+{\rm h.c.} .
\label{h_int_2}
\end{eqnarray}
Here, $\sigma_{+}=\sigma_{-}^{\dagger}=a_{1}^{\dagger}a_{0}$ and we use the shorthand notation $g^{(j)}_k\equiv g_{01,k}^{(jj)}=g_{10,k}^{(j j)*}$, where
\begin{eqnarray*}
g_{lm,k}^{(ij)} = \sqrt{n_{0}}\chi\int dx\varphi^{(j) \dag} _{l}(x) \varphi^{(j)}_{m}(x) \tanh\left(\frac{x -x_i}{\xi }\right) u^{(i)}_{k}.
\end{eqnarray*}
The counter-rotating terms proportional to $b_{k}\sigma^{(j)} _{-}$ and $b^{\dagger}_{k}\sigma^{(j)} _{+}$ that do not conserve the total number of excitations correspond to the intraband terms $(l,m)=(0,0)$ and $(l,m)=(1,1)$, which are ruled out within the RWA. Such an approximation is well justified provided that the emission rate $\gamma$ is much smaller than the qubit transition frequency $\omega_{0}$, as shown in Ref. \cite{muzzamal2017}. \par
\section{Entanglement dynamics} 
\begin{figure}[t!]
\includegraphics[scale=0.7]{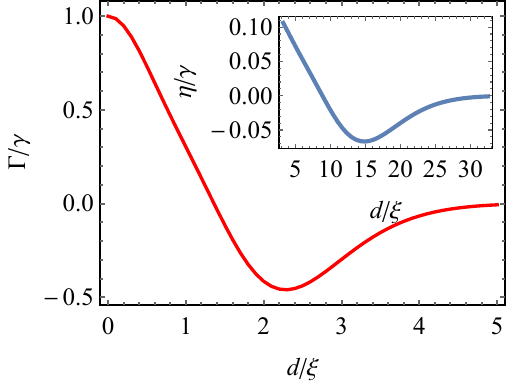}
\caption{(color online)  Collective damping  $\Gamma$ and qubit-qubit interaction parameter $\eta$ (inset) as a function of the soliton separation $d$. We have chosen $\nu=0.75$, for which dark-soliton qubits are well defined.}
\label{fig_damping}
\end{figure}

After tracing over the phonon degrees of freedom \cite{Ficek2002,Lehmberg1970, muzzamal2018}, we obtain the master equation for the two-qubit density matrix $\rho_{q}$
\begin{eqnarray}
\frac{\partial\rho_{\rm q}(t)}{\partial t}  &=& -\frac{i}{\hbar} \left[H_{\rm q},\rho_{\rm q}(t)\right] -{i}\sum^{2}_{i\neq j}\eta_{ij}\left[\sigma_{+}^{i}\sigma_{-}^{j},\rho_{\rm q}(t)\right]\nonumber \\&+&  \sum^{2}_{ij=1}\Gamma_{ij}\left[\sigma_{-}^{j}\rho_{\rm q}(t)\sigma_{+}^{i}-\frac{1}{2} \lbrace \sigma_{+}^{i}\sigma_{-}^{j},\rho_{\rm q}(t) \rbrace \right]  \label{master eq.},
\end{eqnarray}
where
\begin{eqnarray}
\Gamma_{ij} &=& \frac{2L}{\hbar^2}\int_{0}^{\infty}dk g^{(i)}_{k}g_{k}^{(j) \ast} \delta(\omega_{k}-\omega_{0}). \nonumber \\
\eta_{ij} &=&\frac{L}{2\pi \hbar^2}\wp\int_{0}^{\infty}dk g^{(i)}_{k}g_{k}^{(j) \ast}\frac{1}{\left(\omega _{k}-\omega _{0}\right)}\label{parameters},
\end{eqnarray}
and $L$ is the size of the condensate. The diagonal terms $\Gamma_{11}=\Gamma_{22}\equiv\gamma$ are the spontaneous emission rate of each DS-qubit, while the off-diagonal terms $\Gamma_{12}=\Gamma_{21}\equiv \Gamma$ denote the collective damping resulting from the mutual exchange of phonons. The term $\eta_{12}=\eta_{21}\equiv \eta$ represents the phonon-induced coupling between the qubits. Both $\Gamma$ and $\eta$ display a nontrivial dependence on the distance $d$ between the DSs, as depicted in Fig. \ref{fig_damping}. Contrary to what happens for the case of qubits displaced in 1D electromagnetic reservoirs, both parameters vanish for large separations, $d\gg \xi$, rather than displaying a periodic dependence on $d$ \citep{scully_book}. This is a consequence of the local-density approximation (LDA) performed in the computation of the functions $u_k^{(j)}$ and $v_k^{(j)}$, reflecting the local character of the solitons. \par 
\begin{figure}[t!]
\flushleft
\includegraphics[scale=0.48]{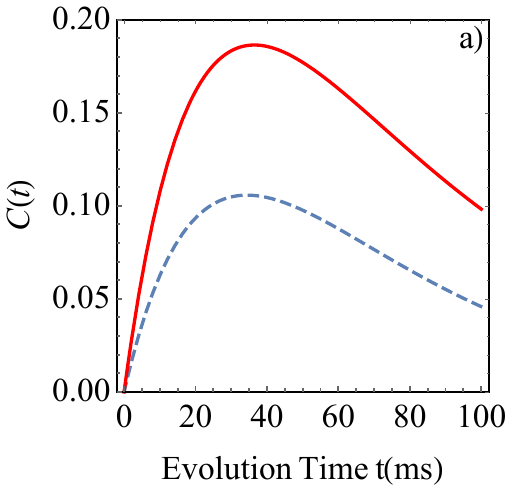}
\includegraphics[scale=0.47]{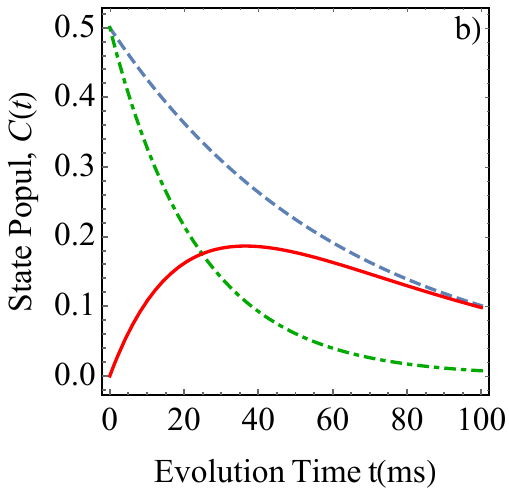}
\caption{(color online) Time evolution of the concurrence $C(t)$ in the absence of driving. Panel a) depicts $C(t)$ for the superposition of maximally entangled Dicke states. $d \simeq \xi$ (dashed curve) and $d \simeq 5\xi/2$ (solid curve). Panel b) shows the population of symmetric state $\left\vert s\right\rangle$ (dashed curve), antisymmetric  state $\left\vert a\right\rangle$ (dotted-dashed curve) and time evolution of concurrence $C(t)$ (solid curve) at distance $d \simeq 5\xi/2$.}     \label{fig_Concurrence}
\end{figure}

We solve Eq. (\ref{master eq.}) in the Dicke basis \cite{Dicke1954}, as shown in Fig. \ref{fig_scheme}b). Depicted are the ground $ \left\vert g\right\rangle=\left\vert g_{1},g_{2}\right\rangle$, the excited $\left\vert e\right\rangle=\left\vert e_{1},e_{2}\right\rangle $, and two intermediate, maximally entangled (symmetric $\left\vert s\right\rangle=\left({\left\vert e_{1},g_{2}\right\rangle+\left\vert g_{1},e_{2}\right\rangle}\right)/{\sqrt{2}}$ and antisymmetric $\left\vert a\right\rangle=\left({\left\vert e_{1},g_{2}\right\rangle-\left\vert g_{1},e_{2}\right\rangle}\right)/{\sqrt{2}})$ states. In this basis, the density matrix elements are given by
\begin{eqnarray}
\rho_{ee}(t)&=& e^{-2\gamma t}\rho_{ee}(0)\nonumber \\
\rho_{ss}(t)&=&e^{-\left(\gamma+\Gamma\right) t}\rho_{ss}(0) \nonumber \\ 
&+& \frac{\left(\gamma+\Gamma\right)}{\left(\gamma-\Gamma\right)}\left(e^{-\left(\gamma+\Gamma\right) t}-e^{-2\gamma t}\right)\rho_{ee}(0)\nonumber \\
\rho_{aa}(t)&=&e^{-\left(\gamma-\Gamma\right) t}\rho_{aa}(0)  \nonumber \\ 
&+& \frac{\left(\gamma-\Gamma\right)}{\left(\gamma+\Gamma\right)}\left(e^{-\left(\gamma-\Gamma\right) t}-e^{-2\gamma t}\right)\rho_{ee}(0)\nonumber \\
\rho_{sa}(t)&=& e^{-\left(\gamma+2i\eta\right) t}\rho_{sa}(0),\label{den. mat ele.} 
\end{eqnarray}
with the condition $\rho_{gg}=1-\rho_{ee}-\rho_{ss}-\rho_{aa}$. The symmetric state $\left\vert s\right\rangle$ is populated, by spontaneous emission, from the state $\left\vert e\right\rangle$ at the superradiant rate $\gamma+\Gamma$, while the anti-symmetric state $\left\vert a\right\rangle$  at the subradiant rate $\gamma-\Gamma$. The quantification of the entanglement is performed by using Wootter's concurrence formula \citep{Wootters1998}, $C(t)=\rm max \lbrace 0,\sqrt{\vartheta}_{1}-\sum^{4}_{n = 2}\sqrt{\vartheta}_{n}\rbrace$, where $\vartheta_{i}$'s denotes the eigenvalues, in the decreasing order, of the hermitian matrix $\zeta=\rho\tilde\rho$. Here, $\tilde\rho= (\sigma_{y}\otimes\sigma_{y})\rho^{\ast}(\sigma_{y}\otimes\sigma_{y})$ describes the spin flip density matrix with $\rho^{\ast}$ and $\sigma_{y}$ being the complex conjugate of $\rho$  and the Pauli matrix, respectively. In the following, we investigate the effect of both $\Gamma$ and $\eta$ in the evolution of $C(t)$ for two different situations: (i) the system is prepared in the state $\left(\left\vert s\right\rangle + \left\vert a\right\rangle\right)/\sqrt{2}$, from which it decays spontaneously, and (ii) the DS-qubits are continuously pumped. In the first case, analytical solutions to Eq. (\ref{den. mat ele.}) provide (see Appendix-\ref{Derivation of Dicke Basis Concurrence})
 \begin{eqnarray}
C(t)=e^{-\gamma t}\sqrt{\sinh^{2}\left(\Gamma t\right)+ \sin^{2}\left(2\eta t\right)}. \label{concurrence}
\end{eqnarray}
Fig. (\ref{fig_Concurrence}) shows $C(t)$ for the initialization of the system in the superposition of maximally entangled states. The concurrence firstly displays a fast increase, being then followed by a slow decay. \par

The time evolution of the initial state that is given by equal populations in the states $\left\vert s\right\rangle$ and $\left\vert a\right\rangle$, i.e. $\rho_{ss}(0)=\rho_{aa}(0)=1/2$, can be seen in panel b) of Fig. (\ref{fig_Concurrence}). It is shown that the decay rate of the state $\left\vert s\right\rangle$ becomes subradiant while the state  $\left\vert a\right\rangle$ decays at the superradiant rate at a sufficiently large distance, $d \simeq 2.5\xi\sim 2-5$ $\mu$m for a BEC in the conditions of \citep{zoran2013}. The concurrence exhibits an appreciably long lifetime ($\sim 80$ ms) due to the asymmetry between the two cascades, eventually reaching the value of the population of the symmetric state $\vert s \rangle$, $C(t)\simeq \rho_{ss}(t)$. 
A major limitation to the concurrence performance could be the DS quantum diffusion, or quantum evaporation \cite{Dziarmaga2004}, a feature that has been theoretically predicted but yet not experimentally validated. Taking into account the latter, a maximum reduction of $20\%$ of the total concurrence lifetime is estimated \cite{muzzamal2017}. In any case, quantum evaporation is expected if important trap anisotropies are present, a limitation that we can overcome with the help of box-like or ring potentials \cite{zoran2013}. Additionally, the effect of the repulsive interaction between two DSs must be considered. Taking the short-range potential described in \cite{tercas}, we estimate a maximum displacement of $\Delta\simeq 0.09d$ for the duration of the concurrence build-up ($\sim 100$ ms, see below), making it unimportant. The numerical simulations on multi-soliton situation found a noticeable displacement for the outer pair of solitons, while the inner 20 solitons stay almost during the lapsed simulation time, $\tau =100$ $\rm ms$ (see Appendix-\ref{The multiple soliton case}).
%
Moreover, the occurrence of impurity condensation on the bottom of  the soliton, due to a sufficiently high concentration of impurities, leads to the breakdown of single particle assumption and spurious qubit energy shift. This can be avoided if fermionic impurities are used instead \cite{gupta2011}. In our numerical estimates, we will consider a very dilute gas of $^{134}$Cs impurities to surround a dense, cigar-shaped $^{85}$Rb condensate, and adjust the parameter $g_{12}$ via Feshbach resonances.  \par
It is worth comparing the entanglement generation protocol presented here with other schemes proposed in the literature, such as plasmon-mediated entanglement in plasmonic waveguides (PW) \cite{Gonzalez2011, Ali2015} and phonon-mediated quantum correlation in nanomechanical resonator \cite{He2017}.  In the case of 1D PWs, a concurrence of lifetime $\sim 8$ ns is obtained at a distance of the order $\sim 600$ nm \cite{Gonzalez2011}. But for transient entanglement mediated by 3D PW, the concurrence lives for a short time ($\sim 4$ ns) \cite{Ali2015}. Here, the concurrence exhibits a substantially large lifetime ($\sim 80$ ms) at much larger distances ($\sim 2-5 $ $\mu$m). Moreover, the investigation of exciton-phonon coupling in hybrid systems (e.g. consisting of semiconductor quantum dots embedded in a nanomechanical resonator) indicates that the stationary concurrence strongly depends on the resonator temperature \cite{He2017}. Fortunately, in our case, thermal effects are negligible (considering BECs operating well below the critical temperature) and therefore the excitations providing the interaction between the DS-qubits (phonons) are purely quantum mechanical in nature. In the present situation, the concurrence is generated due to a considerably large value of the collective damping rate $\Gamma$, as it becomes evident in Fig. (\ref{fig_damping}). \par
\section{Steady-state concurrence with driven DS qubits}
We propose to address the DS qubits with the driving scheme developed in \cite{navon_2016} to excite turbulence in box traps. We use a magnetic field of the form $B(x,t)=B_0+B'\cos(\omega_d t) x$, splitting the impurity $J=1$ manifold. The driving rate is determined by the Rabi frequency $\Omega=g_L\mu_B B' \langle 1 \vert x\vert 0\rangle/\hbar=\mathcal{C}_\alpha g_L\mu_B B'\xi /\hbar $, with $g_L$ denoting the Land\'e factor, $\mu_B$ the Bohr magneton and $\mathcal{C}_\alpha$ being some constant of order $\sim 1$ (Appendix-\ref{Magnetic driving of the qubits}). The inclusion of the driving term modifies the qubit Hamiltonian $H_{\rm q}\rightarrow H_{\rm q}+H_{\rm d}$, where the RWA driving Hamiltonian (obtained for $\omega_d=\omega_0$, for simplicity) reads (Appendix-\ref{Magnetic driving of the qubits})
 \begin{eqnarray}
H_{\rm d}&=&-\hbar\frac{\Omega}{2} \sum^{2}_{j=1} \left[ \sigma^{(j)}_{+} + \sigma^{(j)}_{-}\right] .
\label{driving ham.}
\end{eqnarray}
We solve the master Eq. (\ref{master eq.}) including the driving term in (\ref{driving ham.}) and extract the concurrence $C(t)$ (see Fig. \ref{driven_con.}). Taking $\dot{\rho}_{q}(t)=0$, we obtain the steady-state concurrence (see Appendix-\ref{Derivation of Steady State Concurrence})
 \begin{eqnarray}
C(\infty)= \frac{1}{2}\rm max 	\left\{ 0,\frac{\Omega^2(\gamma \vert U \vert-\Omega^{2})}{\Omega^{4}+\gamma^{2}\left[\Omega^{2}+\frac{1}{4}\lbrace\left(\gamma + \Gamma\right)^{2}+4\eta^{2}\rbrace\right]} \right\}, 
\label{Steady Con}
\end{eqnarray}
where $U=\Gamma+2 i \eta$. As observed, $C(\infty)$ attains its maximum value at the separation $d \simeq 2.5\xi\sim 2-5$ $\mu$m and a Rabi frequency $\Omega\simeq 0.35\gamma$ ($\simeq 5.5$ Hz for our parameters), as shown in Fig. \ref{fig_rabi con.}. This condition is safely met in cold-atom experiments, as magnetic field gradients of $\sim 10$ Gauss/cm allows us to drive the qubits up to $\Omega \sim 1$ kHz (Appendix-\ref{Magnetic driving of the qubits}). The remarkable and appealing feature of DS qubits is the achievement of steady-state concurrence for distances that are much larger than those obtained in other physical systems \cite{Gonzalez2011, Ali2015, He2017}. This paves the stage for unprecedented quantum information applications with phononic platforms. For example, one may think of quantum gates performing at much larger distances than in the case of optical lattices, which achieve logical operations at optical wavelength scales $\sim 800$ nm \cite{Pachos2003}. 
\begin{figure}[t!]
\includegraphics[scale=0.8]{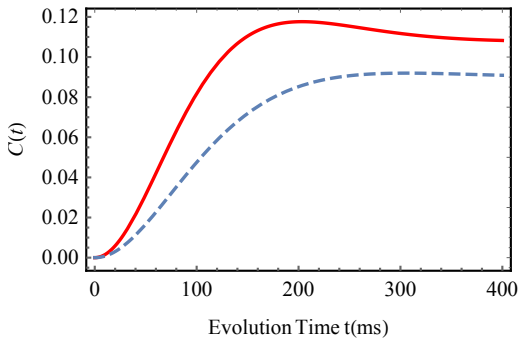}
\caption{(color online) Time evolution of the concurrence $C(t)$ for symmetric pumping ($\Omega_{1}=\Omega_{2}$) at the distance $d= 5\xi/2$. We have chosen $\Omega=0.25\gamma$ (dashed curve) and $\Omega=0.35\gamma$ (solid curve) for illustration.}
\label{driven_con.}
\end{figure}
\begin{figure}[t!]
\includegraphics[scale=0.47]{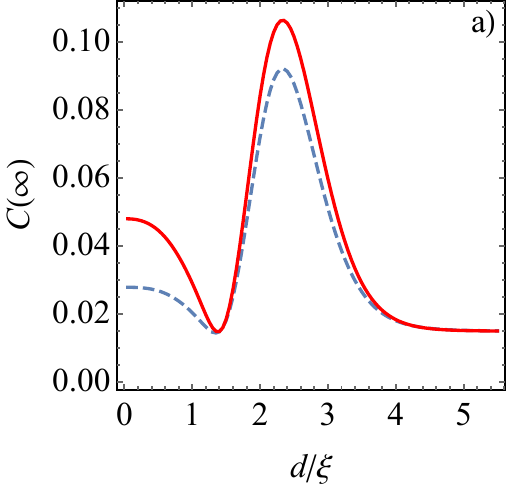}
\includegraphics[scale=0.48]{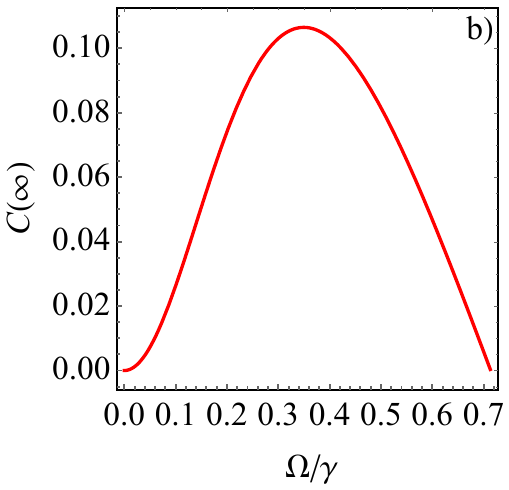}
\caption{(color online) a) The steady-state concurrence $C(\infty)$ as a function of distance $d$ between DS qubits, with $\Omega=0.25\gamma$ (dashed curve) and $\Omega=0.35\gamma$ (solid curve). b) The variation of $C(\infty)$ with the Rabi frequency $\Omega$.}
\label{fig_rabi con.}
\end{figure}
 \par   
 \section{Conclusion}
In conclusion, large-distance entanglement is made possible via the magnetic driving of two dark-soliton qubits, the elements of a recently proposed platform for quantum information processing based solely on matter waves. Dark-soliton qubits consist of two-level systems formed by impurities trapped at the interior of dark solitons, the stable nonlinear depressions produced in quasi one-dimensional Bose-Einstein condensates. The entanglement is mediated by the quantum fluctuations (Bogoliubov excitations, or phonons). Thanks to the large lifetimes of these solitary qubits (being of the order of $100$ ms), an appreciable amount of entanglement can be produced at large distances (a few $\mu$m) for condensates loaded in box potentials. Our conclusion is that dark-soliton qubits are excellent candidates for applications in quantum technologies for which information storage during large times is necessary \cite{Borregaard2015,Jin2017}. We expect that with the development of trapping techniques, allowing for homogeneous condensates of sizes $\sim 100$ $\mu$m, record large-distance pure phononic entanglement $\sim 50 ~\mu$m might be achievable with 10$-$20 dark-solitons, overdoing - or at least matching - the most recent findings with ions \cite{friis_2018}. Also, BECs are good to hybridize with other systems, putting our platform in the run for quantum storage devices with interfaces \citep{tercas_2015, sofia_2016}. 
\appendix 
\section{Bound states in a dark-soliton potential: dark-soliton qubits}
\label{Bound states in a dark-soliton potential: dark-soliton qubits}
\begin{figure}[t!]
\includegraphics[width=0.5\textwidth]{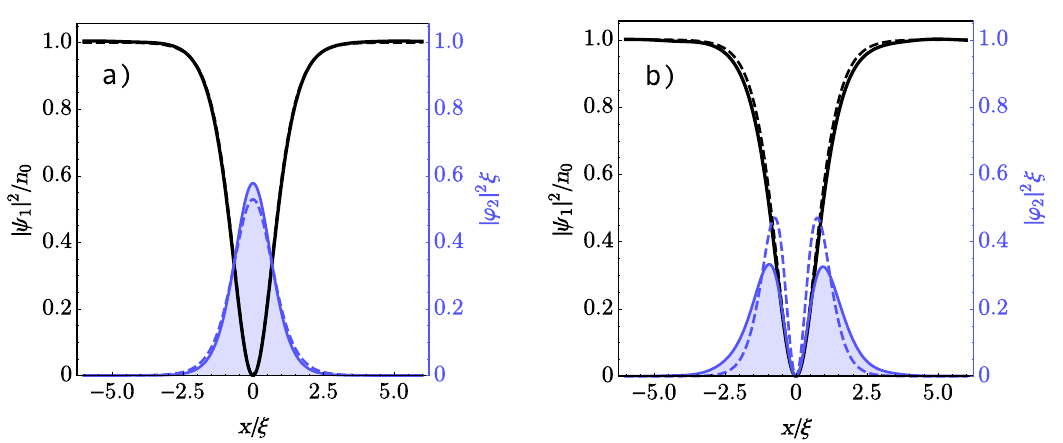}
\caption{(color online) Qubits in a possible experimental situation: Numerical profiles of the dark soliton (black lines) and the impurity eigenstates (blue lines). From left to right, we depict the ground ($\varphi_0(x)$) and the excited ($\varphi_1(x)$) states, respectively, of a fermionic $^{134}$Cs impurity trapped in a $^{85}$Rb BEC dark soliton. The solid lines are the numerical solutions, while the dashed lines are the analytical expression described in the text. We have used the following parameters: $m_\varphi=1.56 m_\psi$, $\chi=0.88 g$ (corresponding to $\nu\simeq 0.75$, as considered in the manuscript). We fix the number of depleted condensate atoms by the dark soliton to be $n_0\xi\simeq 50$, although our numerical simulations (not shown) indicate that the solutions are not very sensitive to its variation. } 
\label{fig_states}
\end{figure}

We consider a dark soliton in a quasi 1D BEC, surrounded by a dilute set of impurities  (a schematic representation can be found in Fig. 1 of the manuscript). The BEC and the impurity particles are described by the wave functions $\psi(x,t)$ and $\varphi(x,t)$, respectively.  At the mean field level, the system is governed by the Gross-Pitaevskii and Schr\"odinger equations, respectively,
\begin{eqnarray}
i\hbar \frac{\partial \psi }{\partial t}&=&-\frac{\hbar ^{2}}{2m_\psi}\frac{
\partial^{2} \psi }{\partial x^{2}} + g \left\vert \psi \right\vert ^{2}\psi
_+\chi \left\vert \varphi\right\vert ^{2}\psi, \nonumber \\ 
i\hbar \frac{\partial \varphi}{\partial t}&=&-\frac{\hbar ^{2}}{2m_\varphi}\frac{
\partial^{2} \varphi}{\partial x^{2}}+\chi\left\vert \psi\right\vert ^{2}\varphi,  \label{gps1}
\end{eqnarray}
The dark solitons are assumed not to be disturbed by the presence of impurities, which we consider to be fermionic in order to avoid condensation at the bottom of the potential. To achieve this, the impurity gas is chosen to be sufficiently dilute, i.e. $\vert \psi\vert^2 \gg \vert\varphi\vert^2 $. Moreover, to decrease the kinetic energy (and therefore increase the effective potential depth), the impurities are chosen to be sufficiently massive. Such a situation can be produced, for example, choosing $^{134}$Cs impurities in a $^{85}$Rb BEC \cite{Michael2015}. Therefore, the impurities can be regarded as free particles that feel the soliton as a potential
\begin{equation}
i\hbar \frac{\partial \varphi}{\partial t}=-\frac{\hbar ^{2}}{2m}\frac{%
\partial^{2} \varphi}{\partial x^{2}}+\chi\left\vert \psi _{\rm sol}\right\vert
^{2}\varphi,  \label{sch. eq. without soliton s}
\end{equation}
where the singular nonlinear solution corresponding to the soliton profile is  $\psi_{\rm sol}(x)=\sqrt{n_{0}}\tanh \left[ x/\xi \right]$. The time-independent version of Eq. (\ref{sch. eq. without soliton s}) reads 
\begin{equation}
(E-\chi n_0)\varphi=-\frac{\hbar ^{2}}{2m_\varphi}\frac{%
\partial^{2} \varphi}{\partial x^{2}}-\chi n_{0}{\rm sech}^{2}\left( \frac{x}{\xi }\right) \varphi,
\label{eq_reflectionless s1}
\end{equation}
To find the analytical solution of Eq. (\ref{eq_reflectionless s1}), the potential is casted in the P\"oschl-Teller form
\begin{equation}
V(x)=-\frac{\hbar ^{2}}{2m\xi ^{2}}\nu (\nu +1){\rm sech}^{2}\left( 
\frac{x}{\xi }\right),  
\label{Potent.}
\end{equation}
with $ \nu=\left(-1+\sqrt{1+4\chi m_\varphi/gm_\psi}\right)/2$. The particular case of $\nu$ being a positive integer belongs to the class of {\it reflectionless} potentials \cite{john07}, for which an incident wave is totally transmitted. For the more general case considered here, the energy spectrum associated to the potential in Eq. (\ref{Potent.}) reads
\begin{equation}
E_{n}^{^{\prime }}=-\frac{\hbar ^{2}}{2m_\varphi\xi ^{2}}(\nu -n)^{2},
\label{energy eigenstates}
\end{equation}
where $n$ is an integer. The number of bound states created by the dark soliton is $n_{\rm bound}=\lfloor \nu+1+\sqrt{\nu(1+\nu)}\rfloor$, where the symbol $\lfloor \cdot\rfloor$ denotes the integer part. As such, the condition for {\it exactly} two bound states (i.e. the condition for the qubit to exist) is obtained if $\nu$ sits in the range
\begin{eqnarray}
\frac{1}{3}\leq \nu < \frac{4}{5},  \label{condition}
\end{eqnarray}
as discussed in the manuscript. At $\nu \geq 4/5$, the number of bound states increases, but this situation is not considered here. In Fig. \ref{fig_states}, we compare the analytical estimates with the full numerical solution of Eqs. (\ref{gps1}), for both the soliton and the qubit wavefunctions, under experimentally feasible conditions.  \\\\

\section{Interaction Hamiltonian}
\label{Interaction Hamiltonian}
As described in the manuscript, the interaction of a system composed of two dark-soliton qubits + quantum fluctuations and impurities can be described by the following many-body Hamiltonian
\begin{equation}
H_{\rm int}=\chi\int dx\Phi ^{\dag }\Psi ^{\dag }\Psi \Phi,
\label{Int. Ham.}
\end{equation}
where $$\Phi(x)=\sum_{l=0}^1\sum_{j=1}^2 \varphi_{l}^{(j)}(x) a_{l}^{(j)}$$ describes the qubit field in terms of the bosonic operators $a_{l}^{(j)}$ annihilating an impurity in the state (or `band') $l=(0,1)$ and soliton $j=(1,2)$. We assume that the potential to be deep enough such that the overlap between the solitons is negligible. Such condition has been verified in additional numerical simulations (not shown here). As such, we use $\varphi_0^{(1)}(x)=\varphi_0^{(2)}(x)\equiv\varphi_0(x)= A_0{\rm sech}  ^{\alpha}(x/\xi)$ and $\varphi_1^{(1)}(x)=\varphi_1^{(2)}(x)\equiv \varphi_1(x)= A_1\tanh  (x /\xi) \varphi_0(x)$, where $A_j (j=0,1)$ are the normalization constants given by
\begin{eqnarray}
A_0&=&\left(\frac{\sqrt{\pi}  \Gamma[\alpha]}{\Gamma[\frac{1+2\alpha}{2}]}\right)^{-\frac{1}{2}}, \nonumber \\
A_1&=&\left(2^{2(1+\alpha)}A_0^2 \left(  \frac{ {_{2}}F_1[\alpha,2(1+\alpha),1+\alpha,-1]}{\alpha} \right.\right.  \nonumber \\   && \left. \left. 
-\frac{ {_{2}}F_1[1+\alpha,2(1+\alpha),2+\alpha,-1]}{1+\alpha}\right.\right.  \nonumber \\   && \left. \left. 
+\frac{ {_{2}}F_1[2+\alpha,2(1+\alpha),3+\alpha,-1]}{2+\alpha}\right)\right)^{-\frac{1}{2}}. \label{normalization}
\end{eqnarray}
Here, 
 $\Gamma[\alpha]$ and ${_{2}}F_1$ represents the Gamma and Hypergeometric function, respectively and $\alpha=\sqrt{2\chi m_\varphi/g m_\psi}$. The inclusion of quantum fluctuations is performed by writing the BEC field as $$\Psi(x)=\left( \psi_{\rm sol}(x)+\sum_{j=1}^2 \delta \psi^{(j)}(x)\right),$$ where $\delta \psi^{(j)}(x)=\sum_k \left(u_k^{(j)}(x) b_k +v^{^{(j)}*}_k(x)b^{\dagger}_k \right)$ and $b_k$ are the bosonic operators verifying the commutation relation $[b_{k},b^{\dagger}_{q}]=\delta_{k,q}$. The amplitudes $u_k^{(j)}(x)$ and $v_k^{(j)}(x)$ satisfy the normalization condition $\vert u_k^{(j)}(x)\vert ^2 -\vert v_k^{(j)}(x)\vert ^2=1$, being, within the local-density approximation (LDA), explicitly given by  \cite{Martinez2011}, 
\begin{eqnarray*}
&&\hspace{-0.5cm}\left. u^{(i)}_{k}(x)= e^{ik(x-x_{i})}\sqrt{\frac{1}{4\pi \xi}}\frac{\mu}{\epsilon_{k}}%
\right. \times \\
&& \hspace{-0.6cm}\left. \left[\left((k\xi)^{2}+\frac{2\epsilon_{k}}{\mu}\right)\left(\frac{k\xi}{2}+i\tanh\left(\frac{x-x_{i}}{\xi}\right)\right)+\frac{k\xi}{%
\cosh^{2}\left(\frac{x-x_{i}}{\xi}\right)}\right] \right.,
\end{eqnarray*}%
and
\begin{eqnarray*}
&&\hspace{-0.5cm}\left. v^{(i)}_{k}(x)= e^{-ik(x-x_{i})}\sqrt{\frac{1}{4\pi \xi}}\frac{\mu}{\epsilon_{k}}%
\right. \times \\
&& \hspace{-0.6cm} \left. \left[\left((k\xi)^{2}-\frac{2\epsilon_{k}}{\mu}\right)\left(\frac{k\xi}{2}+i\tanh\left(\frac{x-x_{i}}{\xi}\right)\right)+\frac{k\xi}{%
\cosh^{2}\left(\frac{x-x_{i}}{\xi}\right)}\right] \right..
\end{eqnarray*}%
%
%
%
%
where $x_j$ is the position of the $j$th soliton. Using the rotating wave approximation (RWA) discussed in the text, the first-order perturbed Hamiltonian can be written as
\begin{eqnarray}
H_{\rm int} &=&\sum_{k}\sum_{i,j=1}^{2}\sum_{l,m=0}^1\left( g^{(ij)}_{lm,k} a^{(i)\dagger}_l a^{(j)}_m b_{k}\right) + {\rm H.c.},.\label{hamilton}
\end{eqnarray}
First, the smallness of the Wannier functions allows us to neglect hopping and, therefore, the cross terms $(i=j)$.

\begin{equation}
g_{lm,k}^{(ij)} = \sqrt{n_{0}}\chi\int dx~\varphi^{(j) \dag} _{l}(x) \varphi^{(j)}_{m}(x) \tanh\left(\frac{x -x_i}{\xi }\right) u^{(i)}_{k}.
\end{equation}
To proceed, we notice that the intraband terms $l=m$ are much smaller than the interband terms $l\neq m$ for the resonant wavevector $k$, i.e. for the phonon mode that is in resonance with the qubit transition $\omega_0$. For illustration, we pick the on-site case (to render the discussion clearer - the off-site coefficients display the same behavior) and compute the intraband terms, whose amplitudes are given by the coefficients $g_{00,k}^{(jj)}\equiv g_{00,k}$ and $g_{11,k}^{(jj)}\equiv g_{11,k}$, with the interband coefficient $g_{10,k}^{(jj)}=g_{10,k}^{(jj)*}\equiv g_{01,k}\equiv g_k$, as illustrated in Fig. \ref{fig_coupling}. As explained in the main text, and as we see below, the validity of our RWA approximation is verified {\it a posteriori}, holding if the corresponding spontaneous emission rate is much smaller than the qubit transition frequency $\omega_0$. Within the present approximation, Eq. (4) of the manuscript is obtained. \\\\ 

\begin{figure}[t!]
\includegraphics[width=0.4\textwidth]{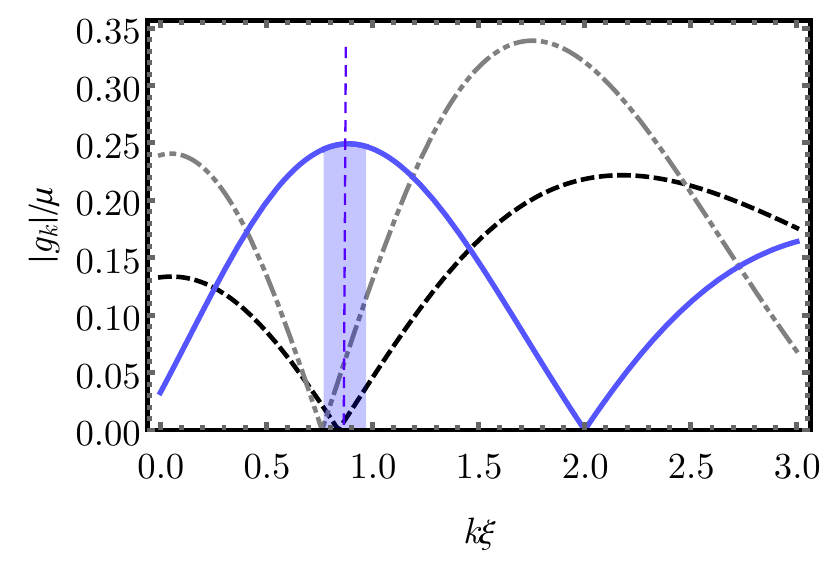}
\caption{(color online) On-site ($i=j$) Interband $g_{01,k}=g_{10,k}^*\equiv g_k$ (solid line) and intraband $g_{00,k}$ and $g_{11,k}$  (dashed and dot-dashed lines, respectively) coupling amplitudes. Near resonance ($k\sim 0.9 \xi^{-1}$), the interband terms clearly dominates over the intraband transitions, allowing us to neglect the latter within the rotating wave approximation,}
\label{fig_coupling}
\end{figure}
\section{Derivation of Dicke Basis Concurrence}
\label{Derivation of Dicke Basis Concurrence}
The computational states of two-two level atoms can be written as product states of individual atoms
\begin{eqnarray}
\vert 1 \rangle &=& \vert e_1 \rangle \otimes  \vert e_2 \rangle, \hspace{0.5cm}   \vert 2 \rangle = \vert e_1 \rangle \otimes  \vert g_2 \rangle, \nonumber \\
\vert 3 \rangle &=& \vert g_1 \rangle \otimes  \vert e_2 \rangle, \hspace{0.5cm}  \vert4 \rangle = \vert g_1 \rangle \otimes  \vert g_2 \rangle.
\end{eqnarray}
The density matrix to calculate the concurrence has the form
\begin{equation}
\rho =\left(                                                   
\begin{array}{cccc}                           
\rho _{11} & 0 &0 & \rho _{14} \\                                  
0 & \rho _{22} & \rho _{23} & 0 \\                                  
0 & \rho _{32} & \rho _{33} &0 \\                                
\rho _{41} & 0 & 0 & \rho _{44}
\end{array}%
\right) ,  \label{Dens}
\end{equation}
for which the square root of the eigenvalues of the matrix $\zeta=\rho \tilde{\rho}$ are
\begin{eqnarray}
\sqrt{\lambda_{1,2}}&=&\sqrt{\rho_{11}\rho_{44}} \pm \vert \rho_{14} \vert, \nonumber \\
\sqrt{\lambda_{3,4}}&=&\sqrt{\rho_{22}\rho_{33}} \pm \vert \rho_{23} \vert. 
\end{eqnarray}
Depending on the largest eigenvalue of the density matrix elements, there are two alternative possibilities to define the concurrence $C={\rm max}\{0,C_1,C_2\}$ with
\begin{eqnarray}
C_1&=&2\left( \vert \rho_{14} \vert -\sqrt{\rho_{22}\rho_{33}} \right), \nonumber \\
C_2&=&2\left( \vert \rho_{23} \vert -\sqrt{\rho_{11}\rho_{44}} \right),
\end{eqnarray}
It is interesting to represent the results of the concurrence in terms of Dicke Basis
\begin{eqnarray}
\vert e \rangle &=& \vert e_1 \rangle \otimes  \vert e_2 \rangle, \hspace{0.5cm}   \vert s \rangle = \frac {1}{\sqrt{2}} \left( \vert e_1 \rangle \otimes  \vert g_2 \rangle + \vert g_1 \rangle \otimes  \vert e_2 \rangle \right), \nonumber \\
\vert g \rangle &=& \vert g_1 \rangle \otimes  \vert g_2 \rangle,  \hspace{0.5cm}
\vert a \rangle = \frac {1}{\sqrt{2}} \left( \vert e_1 \rangle \otimes  \vert g_2 \rangle - \vert g_1 \rangle \otimes  \vert e_2 \rangle \right). \nonumber \\
\end{eqnarray}
for which the matrix to transform original basis to Dicke basis is defined by
\begin{equation}
U= \left(                                                   
\begin{array}{cccc}                           
1 & 0 &0 & 0 \\                                  
0 & \frac {1}{\sqrt{2}} &\frac {1}{\sqrt{2}} & 0 \\                                  
0 & \frac {1}{\sqrt{2}} & -\frac {1}{\sqrt{2}} &0 \\                                
0 & 0 & 0 & 1
\end{array}%
\right) ,  \label{ Trans. Dens. Mat}
\end{equation}
leading to the new density matrix $\rho' =U \rho U^{\dagger}$ with same form as that of Eq. (\ref{Dens}). Dicke basis density matrix elements are related to original density matrix elements as follows,
\begin{eqnarray}
\rho_{ee}&=& \rho_{11},\hspace{1.8cm}\rho_{eg}=\rho_{14}, \nonumber \\
\rho_{gg}&=&\rho_{44},\hspace{1.8cm}\rho_{ge}=\rho_{41}, \nonumber    \\
\rho_{ss}&=&\frac {1}{2}\left(\rho_{22}+\rho_{23}+\rho_{32}+\rho_{33}\right), \nonumber\\
\rho_{sa}&=&\frac {1}{2}\left(\rho_{22}-\rho_{23}+\rho_{32}-\rho_{33}\right), \nonumber \\
\rho_{aa}&=&\frac {1}{2}\left(\rho_{22}-\rho_{23}-\rho_{32}+\rho_{33}\right), \nonumber\\ 
\rho_{as}&=&\frac {1}{2}\left(\rho_{22}+\rho_{23}-\rho_{32}-\rho_{33}\right). 
\label{elements}
\end{eqnarray}
In the Dicke Basis, the eigenvalues of the matrix $\zeta=\rho \tilde{\rho}$ are
\begin{eqnarray}
\sqrt{\lambda_{1,2}}&=&\sqrt{\rho_{ee}\rho_{gg}} \pm \vert \rho_{eg} \vert, \nonumber \\
\sqrt{\lambda_{3,4}}&=&\frac {1}{2} \left(\sqrt{\left(\rho_{ss}+\rho_{aa}\right)^2-\left(\rho_{sa}+\rho_{as}\right)^2} \right. \nonumber \\ 
 & \pm &\left. \sqrt{\left(\rho_{ss}-\rho_{aa}\right)^2-\left(\rho_{sa}-\rho_{as}\right)^2}\right).
\end{eqnarray}
Therefore, the alternative form of the concurrence becomes
\begin{eqnarray}
C_1&=&2\vert \rho_{eg} \vert - \sqrt{\left(\rho_{ss}+\rho_{aa}\right)^2-\left(\rho_{sa}+\rho_{as}\right)^2}, \nonumber \\
C_2&=&\sqrt{\left(\rho_{ss}-\rho_{aa}\right)^2-\left(\rho_{sa}-\rho_{as}\right)^2} -2\sqrt{\rho_{ee}\rho_{gg}}. \nonumber \\
\label{comb. of conc.}
\end{eqnarray}
Let the system is prepared, initially, in the state $\left(\vert s \rangle + \vert a \rangle\right)/\sqrt{2}$ and using the density matrix elements of Eq. (\ref{elements}), the concurrence can be written as
\begin{eqnarray}
C=e^{-\gamma t}\sqrt{{\rm sinh^2(\Gamma t)}+{\rm sin^2(2\eta t)}},
\end{eqnarray}
which is the required proof.
\section{The multiple soliton case}
\label{The multiple soliton case}
Although not necessary to the understanding of the present case, we have performed numerical simulations on multi-soliton situation. One of the main concerns in related to their mutual repulsion. In a box potential of size $L\simeq 100$ $\mu$m, we can imprint over 20 solitons, separated by distance of $d=2.5 ¤\xi$, as in the main text. 
\begin{figure}[h!]
\includegraphics[width=0.35\textwidth]{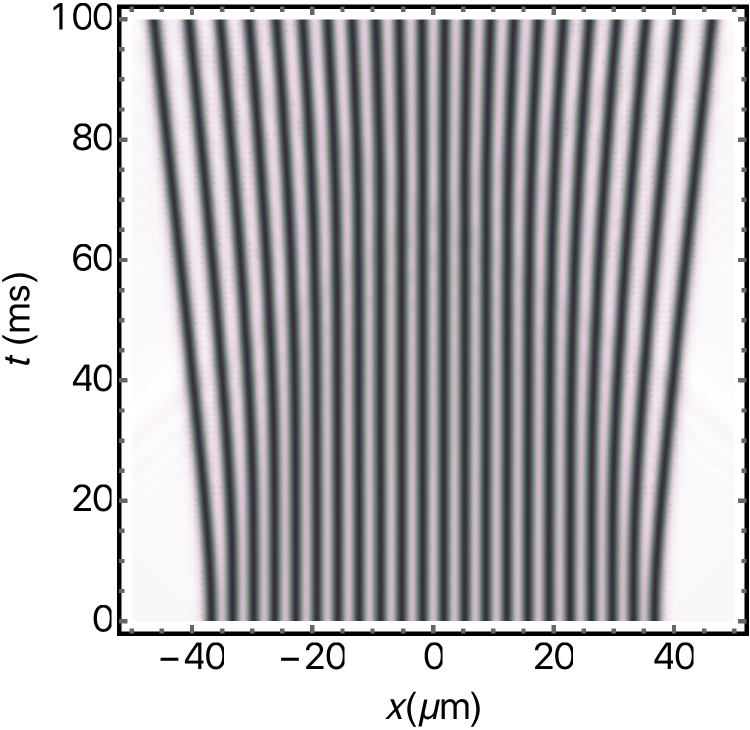}
\caption{(color online) A box potential of size $L=100$ $\mu$m containing 24 solitons. Noticeable displacement is only found for the outer pair of solitons, while the inner 20 solitons stay almost during the lapsed simulation time, $\tau =100$ $ms$, larger than the concurrence build-up time of $\sim 80$ ms described in the manuscript.}
\label{fig_solitons}
\end{figure}
As we can see from Fig. \ref{fig_solitons}, their mutual repulsion is very small, an therefore deterioration of the entanglement is expected to be negligible within the concurrence build-up time ($\sim 80$ ms, as described in the manuscript). Solving the master equation for the multi-soliton case is extremely demanding computationally, and will therefore be addressed in a separate publication.

\section{Magnetic driving of the qubits}
\label{Magnetic driving of the qubits}
In order to attain a finite steady-state concurrence in a pair of dark-soliton qubits, we must drive the transition $\vert 0 \rangle \leftrightarrow \vert 1 \rangle$ with a cw field. In atoms and ions, this is simply performed with an external laser, which couples to the electronic transitions ($\omega_0 \sim 10^{14}$ Hz) via a dipole term $\sim \mathbf{p}\cdot \mathbf{A}$, with an amplitude given by the Rabi frequency of $\Omega=\mathbf{p}\cdot \mathbf{A}/\hbar $. Here, we are dealing with transitions involving the center-of-mass motion of the impurities, for which the typical frequencies are of the same order of the chemical potential of the BEC, $\omega_0\sim\mu/\hbar \sim$ kHz. A possible way to access this transition is by applying a time-varying magnetic field gradient along the BEC axis, $\mathbf{B}(x,t)=(B_0+B'e^{i\omega_d t}x) e_{\bf x}$. This allows to Zeeman split the impurity $J=1$ manifold, which results in a driving Hamiltonian of the form  
\begin{equation}
H_{\rm drive}=-\bm \mu \cdot {\bf B}=-\int dx ~\varphi(x)^\dagger g_L\mu_B B(x)\varphi(x). 
\end{equation}
By using the decomposition into the states $\vert 0 \rangle$ and $\vert 1 \rangle$ discussed above, we can re-write the driving Hamiltonian as
\begin{eqnarray}
H_{\rm drive} &=&-\hbar \frac{\Omega}{2} \sum_{i=1}^2 \left(e^{i\omega_d t}\sigma_+^{i}+\sigma_-^{i}e^{-i\omega_d t}\right) \nonumber \\
&+&E_{\rm Zeeman}(a_1^\dagger a_1+a_0^\dagger a_0),
\end{eqnarray}
where $E_{\rm Zeeman}=g_L \mu_B B_0$ is a Zeeman shift that we can absorb in the definition of $\omega_0$ (in practice, by choosing a quadrupolar field configuration - as in the case of a magnetic field produced by anti-Helmholtz coils, we can safely assume $B_0\sim 0$), and $\Omega=g_L \mu_B B'\langle 1 \vert x \vert 0 \rangle/\hbar$ is the Rabi frequency, which explicitly reads
\begin{equation}
\Omega = \frac{\mathcal{C_\alpha}}{\hbar}g_L\mu_B B'\xi,
\end{equation}
where $\mathcal{C}_\alpha =\int \varphi_1(x) x\varphi_0(x)$ is a constant of the order of unit ($0.6\leq \mathcal{C}_\alpha\leq 0.86$ for $0.5\leq \alpha \leq 2.0$). A magnetic field gradient of the order $\sim 10$ Gauss/cm is currently produced in cold atom experiments, allowing us to attain a Rabi frequency up to $\Omega \sim 150$ Hz, around $10\%$ of the qubit transition energy $\omega_0$. The latter fairly exceeds the requirements for a maximum concurrence situation, achieved for $\Omega\simeq 0.35 \gamma \simeq 5.5 $ Hz for the conditions of the numerical examples discussed in the manuscript (see Ref. \cite{muzzamal2017} for details on the relation between $\gamma$ and $\omega_0$). \\\\
\section{Derivation of Steady State Concurrence}
\label{Derivation of Steady State Concurrence}
 To find the steady state concurrence, Eq. (\ref{master eq.}) can be written as
\begin{eqnarray}
 \frac{i}{\hbar} \left[H_{\Omega},\rho_{q}\right]+\sum^{2}_{i\neq j}\eta_{ij}\left[\sigma_{+}^{i}\sigma_{-}^{j},\rho_{q}\right] = \nonumber \\
 \sum^{2}_{ij=1}\Gamma_{ij}\left[\sigma_{-}^{j}\rho_{q}\sigma_{+}^{i}\right.
 - \left.\frac{1}{2} \lbrace \sigma_{+}^{i}\sigma_{-}^{j},\rho_{q} \rbrace \right], 
 \label{final master eq.11}
\end{eqnarray}
with the density matrix elements
\begin{eqnarray}
 \rho_{ee}&=&\frac{\Omega^4}{\gamma^2\left( \left( \gamma+\Gamma\right)^2+4\left(\eta^2 +\Omega^2\right) \right)+4\Omega^4}, \nonumber \\
  \rho_{ss}&=&\frac{\Omega^2\left(2\gamma^2+\Omega^2 \right)}{\gamma^2\left( \left( \gamma+\Gamma\right)^2+4\left(\eta^2 +\Omega^2\right) \right)+4\Omega^4}, \nonumber \\
   \rho_{aa}&=&\frac{\Omega^4}{\gamma^2\left( \left( \gamma+\Gamma\right)^2+4\left(\eta^2 +\Omega^2\right) \right)+4\Omega^4}, \nonumber \\
    \rho_{gg}&=&\frac{\gamma^2\left( \left( \gamma+\Gamma\right)^2+2\left(2\eta^2 +\Omega^2\right) \right)+\Omega^4}{\gamma^2\left( \left( \gamma+\Gamma\right)^2+4\left(\eta^2 +\Omega^2\right) \right)+4\Omega^4}, \nonumber \\
     \rho_{ge}&=&-\frac{\gamma\left(\gamma+\Gamma+2i\eta \right)\Omega^2}{\gamma^2\left( \left( \gamma+\Gamma\right)^2+4\left(\eta^2 +\Omega^2\right) \right)+4\Omega^4}, \nonumber \\
  \rho_{es}&=&\frac{i \sqrt{2} \gamma \Omega^3}{\gamma^2\left( \left( \gamma+\Gamma\right)^2+4\left(\eta^2 +\Omega^2\right) \right)+4\Omega^4}, \nonumber \\
   \rho_{gs}&=&\frac{i \sqrt{2} \gamma \Omega\left(\gamma\left(\gamma+\Gamma+2i\eta \right)+\Omega^2\right)}{\gamma^2\left( \left( \gamma+\Gamma\right)^2+4\left(\eta^2 +\Omega^2\right) \right)+4\Omega^4},
\end{eqnarray}
where, $\rho_{ij}=\rho_{ji}^*$ and all other density matrix elements are zero. Here, we assume a symmetric pumping for which $\Omega_1=\Omega_2$.
Using Wootter's criteria to find the concurrence and simplified expressions of the density matrix elements, we obtained
 \begin{eqnarray}
C(\infty)= \frac{1}{2}\rm max 	\left\{ 0,\frac{\Omega^2(\gamma \vert U \vert-\Omega^{2})}{\Omega^{4}+\gamma^{2}\left[\Omega^{2}+\frac{1}{4}\lbrace\left(\gamma + \Gamma\right)^{2}+4\eta^{2}\rbrace\right]} \right\},    \nonumber \\
\label{st. st. con.}
\end{eqnarray}
where $U=\Gamma+2 i \eta$. Eq. (\ref{st. st. con.}) is the final expression of the steady state concurrence.


\section*{Acknowledgements}
We thank Raphael Lopes and Sofia Ribeiro for stimulating discussions. This work is supported by the IET under the A F Harvey Engineering Research Prize, FCT/MEC through national funds and by FEDER-PT2020 partnership agreement under the project UID/EEA/50008/2019. The authors also acknowledge the support from Funda\c{c}\~{a}o para a Ci\^{e}ncia e a Tecnologia (FCT-Portugal), namely through the grants No. SFRH/PD/BD/113650/2015 and No. IF/00433/2015. E.V.C. acknowledges partial support from FCT-Portugal through Grant
No. UID/CTM/04540/2013.










\end{document}